\documentclass[aps,twocolumn,prl]{revtex4}
\usepackage{graphicx}% Include figure files
\usepackage{dcolumn}% Align table columns on decimal point
\usepackage{bm}% bold math
\usepackage{amsmath}

\begin{document}
\preprint{APS/123-QED}

\title{Zero-field Quantum Critical Point in CeCoIn$_5$}

\author{Y. Tokiwa$^1$}
\author{E. D. Bauer$^2$}
\author{P. Gegenwart$^1$}

\affiliation{$^1$I. Physikalisches Institut,
Georg-August-Universit\"{a}t G\"{o}ttingen, 37077 G\"{o}ttingen,
Germany}
\affiliation{$^2$Los Alamos National Laboratory, Los Alamos, New Mexico 87545, USA}

\date{\today}% It is always \today, today,
             %  but any date may be explicitly specified
\pacs{}% PACS, the Physics and Astronomy
                             % Classification Scheme.
%\keywords{Suggested keywords}%Use showkeys class option if keyword
                              %display desired

\begin{abstract}
Quantum criticality in the normal and superconducting state of the heavy-fermion metal CeCoIn$_5$ is studied by measurements of the magnetic Gr\"{u}neisen ratio, $\Gamma_H$, and specific heat in different field orientations and temperatures down to 50~mK. Universal temperature over magnetic field scaling of $\Gamma_H$ in the normal state indicates a hidden quantum critical point at zero field. Within the superconducting state the quasiparticle entropy at constant temperature increases upon reducing the field towards zero, providing additional evidence for zero-field quantum criticality.
\end{abstract}

\maketitle

The interplay between magnetism and unconventional superconductivity is one of the central issues in condensed matter physics. Several material classes such as cuprates, iron-pnictides or heavy-fermion metals display non-Fermi liquid (NFL) normal-state properties that may arise from a quantum critical point (QCP) at which a long-range ordered phase is continuously suppressed to zero temperature~\cite{broun-nphys08,Abrahams-JPhys11,Pfleiderer-RMP09}. However, NFL properties in the vicinity of such hidden QCPs cannot be investigated without destroying the superconducting (SC) state by sufficiently large magnetic fields, which also strongly influence quantum criticality~\cite{Sachdev-pss10,Cuprate_NFL_Hifield}. On the other hand, if there is universal scaling of the free energy with respect to temperature and some control parameter in the normal state it is possible to prove the existence and characterize the nature of the hidden QCP. Most sensitive probes of such scaling behaviors are the Gr\"uneisen ratios $\Gamma_r=T^{-1}(dT/dr)_S$ (r: pressure of magnetic field, $S$: entropy), which diverge in the approach of the QCP~\cite{zhu}. Since magnetic field can easily be varied in-situ and the magnetic Gr\"uneisen ratio which equals the adiabatic magnetocaloric effect is directly measurable with high precision, a field-tuned QCP hidden by superconductivity as proposed for the heavy-fermion metal CeCoIn$_5$~\cite{paglione:prl-03,bianchi:prl-03b,Donath-prl08,Ronning-prb05,Zaum-prl11,curro-prl03} or Ce$_2$PdIn$_8$~\cite{tokiwa-prb11,Dong-PRX11} should best be characterized by the latter property. 

%\cite{mathur:nature-98,Hashimoto-science12}. ~\cite{LohneysenHilbertV:Ferimq}. . 

CeCoIn$_5$ undergoes a SC transition at $T_c$=2.3\,K, which is the highest among the ambient-pressure heavy-fermion superconductors~\cite{petrovic:jpcm-01}, but low enough to neglect phononic contributions to heat capacity and suppress the SC state by moderate magnetic fields. CeCoIn$_5$ has attracted considerable attention as very clean metal close to long-range magnetic order, which displays intriguing SC and normal properties~\cite{movshovich:prl-01,bianchi:prl-03a,paglione:prl-03,bianchi:prl-03b,Donath-prl08,Ronning-prb05,Zaum-prl11,curro-prl03}. Electrical resistivity, specific heat and thermal expansion display NFL behavior in the normal state above $T_c$ which extend to mK temperatures close to the SC upper critical field $H_{c2}$~\cite{paglione:prl-03,bianchi:prl-03b,sidorov:prl-02,Donath-prl08,Ronning-prb05,Zaum-prl11,curro-prl03}. On the other hand, for fields sufficiently larger than $H_{c2}$ a crossover to Fermi liquid (FL) behavior has been recovered, which allows to extrapolate to a field-induced QCP. 
Early electrical resistivity~\cite{paglione:prl-03} and specific heat~\cite{bianchi:prl-03b} measurements suggest a field-induced QCP very close to the upper critical field, which amounts to $H_{\rm c2}$=5\,T and 12\,T for the field along and perpendicular to [001], respectively. Subsequent Hall effect~\cite{Singh-prl07} and thermal expansion~\cite{Zaum-prl11} measurements, however, have extrapolated the critical field $H_c$ for the field-tuned QCP to values clearly below $H_{c2}$, i.e. around 4\,T for $H\parallel$ [001]. A peculiar observation is the dependence of estimated $H_{c}$ on the current direction of electrical resistivity for $H\parallel$[001]. The $A$-coefficient of FL resistivity, $\rho$=$\rho_0+AT^2$, measured with the current along basal plane diverges towards 5\,T~\cite{paglione:prl-03,Tanatar-science07,Ronning-prb06}, while $A$ with the current along [001] indicates a significantly lower field for the singularity, 1.5-3T~\cite{Malinowski-prb05}. In view of the controversy concerning the exact location of a possible field-induced QCP in CeCoIn$_5$, systematic studies of the magnetocaloric effect, which is the most sensitive thermodynamic probe of field-tuned quantum criticality are highly desirable. 

Below, we report a systematic investigation of the temperature, field and field-angle dependence of the magnetic Gr\"uneisen parameter in the normal state of CeCoIn$_5$. Surprisingly, we have discovered universal quantum critical scaling indicating a zero-field QCP. This is further supported by an enhanced quasiparticle entropy, derived from the magnetic Gr\"uneisen ratio and specific heat within the SC state in the vicinity of zero field. This implies CeCoIn$_5$ is exceptional as clean material at a QCP without additional fine tuning of composition, pressure or magnetic field.

High quality single crystals were grown by the self-flux method. The specific heat $C(T,H)$ and magnetic Gr\"uneisen ratio $\Gamma_H=T^{-1}(dT/dH)_S$ were measured with very high resolution in a dilution refrigerator with a SC magnet equipped with an additional modulation coil by utilizing heat-pulse and alternating field techniques, respectively, as described in~\cite{tokiwa-rsi11}. The magnetic field has been applied along four different field angles between the [001] and [100] direction.

\begin{figure}[htb]
\includegraphics[width=\linewidth,keepaspectratio]{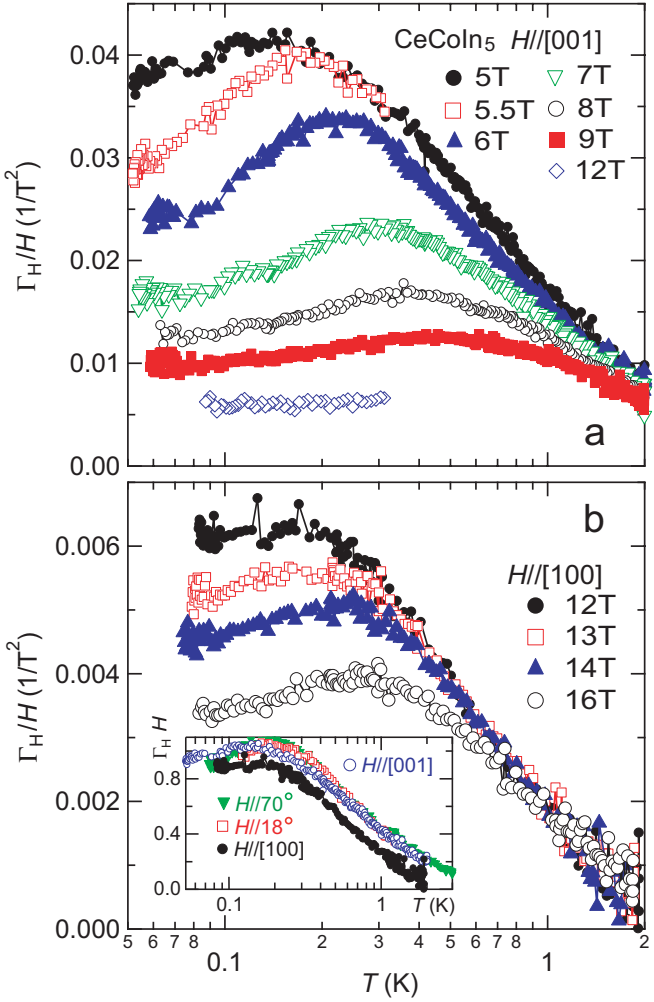}
\caption{(Color online) Magnetic Gr\"uneisen ratio divided by magnetic field, $\Gamma_{\rm H}/H$, of CeCoIn$_5$ in the normal conducting state plotted against temperature for fields applied along [001] (a) and [100] (b). The inset displays $\Gamma_{\rm H}H$ as a function of temperature for different field angles close to the respective upper critical fields. Labels 18$^{\circ}$ and 70$^{\circ}$ denote field angles from [100] towards [001]. The applied magnetic field is 5, 6, 10 and 12\,T for $H\parallel$[001], 70$^{\circ}$, 18$^{\circ}$ and parallel [100], respectively.}
\end{figure}

We first focus on the magnetic Gr\"uneisen ratio in the normal state at various fields and field orientations, cf. Fig. 1. Upon cooling from high temperatures, $\Gamma_{\rm H}(T)/H$ first increases until it passes a maximum and, as most clearly seen for fields above 6~T, saturates at lowest temperatures. Such temperature dependence is characteristic for the crossover between NFL behavior at high $T$ and a FL state at low $T$~\cite{GarstM:SigctG}, which e.g. at 5~T occurs near 0.14~K. The data are thus incompatible with a QCP at $H_{c2}$.  As shown in Fig. 1b, similar behavior is also found for $H\parallel [100]$ and all intermediate field orientations. Within the quantum critical regime, $\Gamma_{\rm H}(T)$ is expected to display a power-law divergence upon cooling. However, for fields $H>H_{c2}$, we only observe an almost linear increase on semi-log scale upon cooling. This indicates, that the QCP must be far below $H_{c2}$. 
%Furthermore, when the system is tuned to quantum critical field, power-law divergence of $\Gamma_{\rm H}$ gives rise to large $\Gamma_{\rm H}$ at low temperatures and it reaches up to $\sim$10\,/T at 0.1\,K in YbRh$_2$Si$_2$~\cite{tokiwa-prl09}. The much smaller $\Gamma_{\rm H}$ ($\sim$0.2\,/T at maximum) also points towards the quantum critical field in CeCoIn$_5$ located rather far away from $H_{c2}$. 
Further information on the critical field $H_c$ of the QCP can be obtained by analyzing the magnetic Gr\"uneisen ratio in the FL state at $T\rightarrow 0$. For a field-induced QCP which follows universal scaling it is expected that $\Gamma_{\rm H}(H-H_c)=-\nu(d-z)$, where $d$ is spatial dimension, $\nu$ the correlation length exponent and $z$ the dynamical critical exponent~\cite{zhu,GarstM:SigctG}. Thus, if the data follow such behavior, we may determine $H_c$ and obtain important information on the quantum critical exponents. For the analysis, we include data for fields parallel and perpendicular to the [001] direction, as well as, for four different intermediate field directions at the respective upper critical fields (cf. inset of Fig. 1b). The overall temperature dependencies for all these field directions are similar and $\Gamma_{\rm H}H$ approaches a common value for $T\rightarrow 0$. As shown below, this is a consequence of universal quantum critical scaling, since the above prefactor, $-\nu(d-z)$, characterizes the nature of quantum criticality and is independent of the direction of applied field. Furthermore, it indicates a quantum critical field $H_c$ being close to zero.

The $T\rightarrow 0$ values for $\Gamma_{\rm H}$ in the FL regime at various different field values and orientations are plotted versus inverse field in the inset of Fig. 2. Remarkably, all data points collapse on a universal line through the origin, regardless of the field direction. Given the sizable magnetic anisotropy of this system~\cite{tayama:prb-02}, this isotropic divergence towards zero field with a common prefactor ($\Gamma_{\rm H}H\approx$0.82 for $T\rightarrow$0) provides strong evidence for universal quantum critical scaling with $H_c$ close to zero.

Following the theory of~\cite{zhu,GarstM:SigctG,LohneysenHilbertV:Ferimq}, we assume that critical behavior is governed by a single diverging time scale near the QCP. The critical contribution to the free energy, $F_{cr}$ for magnetic field as tuning parameter can then be expressed by

\begin{equation}
F_{cr}=aT^{(d+z)/z}\phi\left(\cfrac{bh}{T^{1/\nu z}}\right), 
\end{equation}

where $h=H-H_c$. $a$ and $b$ are non-universal constants. The thermodynamic properties are then expressed by derivative(s) of the free energy, and therefore should collapse on a scaling function of $T/h^{\nu z}$. Remarkably all our $\Gamma_{\rm H}$ data collapse on a single curve in a log-log scaling plot of the form $\Gamma_{\rm H}H$ vs. $T/H^{3/2}$ (Fig.~2), indicating that $h=H$ equivalent to $H_c=0$ and implying $\nu z$=3/2. Within the FL regime at low temperatures and $H>0$ the data scatter around a constant value of $\Gamma_{\rm H}H\approx$0.82 (cf. also the slope of line in the inset of Fig. 2), whose meaning is addressed below. We first address the universal behavior within the NFL regime at large $T/H^{3/2}$, for which we observe $\Gamma_{\rm H}H \propto (T/H^{3/2})^{-4/3}$.

In equation~(1) for the critical contribution of the free energy, we expand the dimensionless argument $y=(bh)/T^{1/\nu z}$ of the scaling function $\phi$ for high temperatures, $y\ll 1$, according to $\phi(y)\approx \phi(0)+1/2\phi''(0)y^2$. Note here that $y$-linear term vanishes, because it corresponds to $H$-linear dependence of free energy and would yield a spontaneous magnetization. The magnetic Gr\"{u}neisen ratio is derived from the free energy by calculating $\Gamma_{\rm H}=[\partial^2F_{cr}/(\partial T\partial H)]/[T\partial^2F_{cr}/\partial T^2]$, yielding $\Gamma_{\rm H}H\sim [b'T^{2/\nu z}/H^2+1]^{-1}$. Here $b'$ is another non-universal constant. The expansion for high temperatures then yields $\Gamma_{\rm H}H\sim (T/H^{\nu z})^{-2/\nu z}$. The experimentally observed high-temperature exponent, $-4/3=-2/\nu z$ (cf. dashed black line in Fig. 2), is perfectly consistent with $\nu z=3/2$, obtained from the argument of scaling function, $\phi(T/H^{3/2})$. Previously, it has been shown that within the NFL regime the "thermal" Gr\"uneisen parameter given by the ratio of thermal expansion to specific heat diverges like $T^{-2/3}$ for CeCoIn$_5$~\cite{Donath-prl08}. The quantum critical scaling predicts that this exponent equals $-1/(\nu z)$~\cite{zhu,GarstM:SigctG}. Thus, $\nu z=3/2$ is fully consistent with the previous thermal expansion study.

\begin{figure}[t]
\includegraphics[width=\linewidth,keepaspectratio]{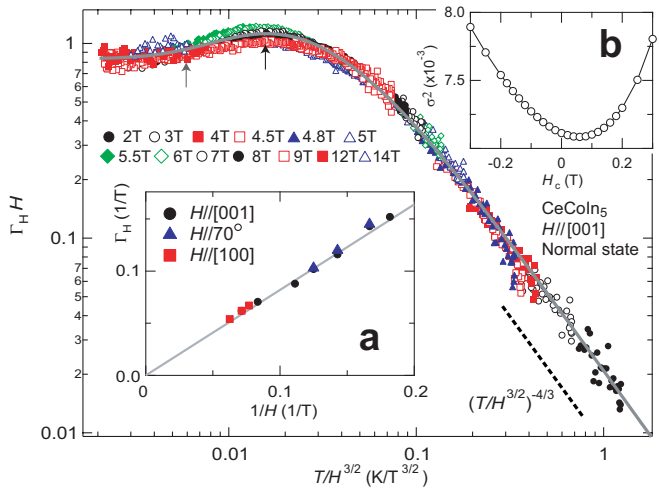}
\caption{(Color online) Magnetic Gr\"uneisen ratio of CeCoIn$_5$ for the field along [001] as $\Gamma_{\rm H}H$ vs $T/H^{3/2}$, on double logarithmic scale. The grey solid line indicates a phenomenological fit by the function $f(x)=c_1/(x^{4/3}+1)-c_2/(x_0+x^2)$, where $x=b'T/H^{3/2}$, $c_1$, $c_2$, $b'$ and $x_0$ are free parameters. The positions of the maximum and inflexion point, respectively 0.015 and 0.006\,KT$^{-3/2}$, that define the crossover between the NFL and FL regime, are indicated by arrows. Inset (a) shows $\Gamma_{\rm H}(T\rightarrow 0)$ values in the FL regime versus the inverse magnetic field for different field directions. Grey solid line indicates linear behavior 0.82/$H$. Inset (b) shows the weighted mean-square deviation from the phenomenological function, $f(x)$, versus the quantum critical field, $H_c$.}
\end{figure}

A further constraint on the critical exponents characterizing quantum criticality in CeCoIn$_5$ is obtained from the magnetic Gr\"uneisen behavior in the FL state following $\Gamma_{\rm H}H\approx$0.82. Since scaling predicts $\Gamma_{\rm H}=-\nu(d-z)/(H-H_c)$ ~\cite{zhu,GarstM:SigctG}, i.e. $-\nu(d-z)\approx 0.82$ it follows that $\nu d\approx 0.68$ which provides a strong constraint for the nature of quantum criticality in this system: assuming a dimensionality of the critical fluctuations in the NFL regime of $d=2$~\cite{Donath-prl08}, the critical correlation length exponent $\nu\approx 1/3$, which is different from the standard Hertz-Millis-Moriya type of quantum criticality for which $\nu=1/2$~\cite{LohneysenHilbertV:Ferimq}.

\begin{figure}[t]
\includegraphics[width=\linewidth,keepaspectratio]{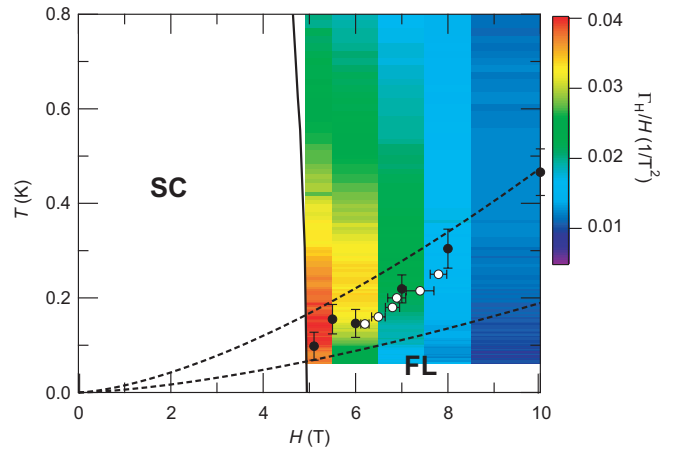}
\caption{(Color online) $T$-$H$ phase diagram of CeCoIn$_5$ for magnetic fields applied along [001]. Color code in the normal conducting state represents $\Gamma_H/H$. Onset of FL behavior determined by thermal expansion~\cite{Zaum-prl11} (solid circles) and Hall effect~\cite{Singh-prl07} (open circles) measurements is included. The two broken lines represent the cross-over between the NFL and FL regions, $T_{\rm FL}$=0.015\,KT$^{2/3}H^{3/2}$ and $T_{\rm FL}$=0.006\,KT$^{2/3}H^{3/2}$, as defined by the maximum position and inflexion point in $\Gamma_HH$, respectively (see arrows in Fig.~2.).}
\end{figure}

In order to examine how close the quantum critical field ($H_c$) is to zero, we investigated the optimum collapse of the data in the scaling plot by tiny variation of $H_c$. The phenomenological function $f(x)=c_1/(x^{4/3}+1)-c_2/(x_0+x^2)$, shown in Fig.~2, fits very well the experimental data for the whole parameter range. Here, $x=b'T/(H-H_c)^{3/2}$, $c_1$, $c_2$, $x_0$ and $b'$ are fitting parameters ($H_c$=0 in Fig.~2). The first term in $f$ represents the expected behavior for $x\gg1$ (see above), while the second one is a phenomenological correction for low $x$ region. We calculated the weighted mean-square deviation ($\sigma^2$) of the data from the phenomenological function as a function of $H_c$, where $\sigma^2$ is defined as $\sigma^2=\sum_i^N(1/N)[\Gamma_{\rm H}(x_i)h_i-f(x_i)]^2/f(x_i)^2$ ($i$ refers to each data point, $h_i=H_i-H_c$, $x_i=T_i/h_i^{3/2}$ and $N$ is the number of data points). Since simple non-weighted mean square deviation will be dominated by the data points at low temperatures only, where $\Gamma_{\rm H}h$ is large, we weighted it with $1/f(x_i)^2$, such that each data point contributes equally. For each calculated $H_c$ value from $-0.3$ to 0.3\,T, the magnetic Gr\"uneisen ratio data have been fitted by $f(x)$ and a respective $\sigma^2$ value has been estimated. The optimum collapse of the data with minimal $\sigma^2$ is obtained for $H_c$=0.06\,T. Importantly, the main contribution to $\sigma^2$ arises from the scattering of the data, while the quality of the collapse is subleading for small values of the critical field. In particular, the difference between the $\sigma^2$ values for 0.06~T and zero field is marginal, which justifies to conclude zero-field quantum criticality in the system.
%consists of two contributions; one from the quality of collapse and the other from experimental scatter. Since $\Gamma_{\rm H}$ values ($\sim$0.2\,/T at maximum) are very small, compared to the other heavy fermion materials (up to $\Gamma_{\rm H}\sim$10\,/T~\cite{tokiwa-prl09,gegenwart-jlt10}), the contribution from the scatter is relatively large and is estimated to be $\sim$2$\times$10$^{-3}$, about one third of the minimum $\sigma^2$. Therefore, difference between $\sigma^2$ values for $H_c$=0\,T and 0.06\,T (3$\times$10$^{-5}$) is marginal and the quantum critical field is virtually zero within the experimental accuracy.

\begin{figure}[t]
\includegraphics[width=\linewidth,keepaspectratio]{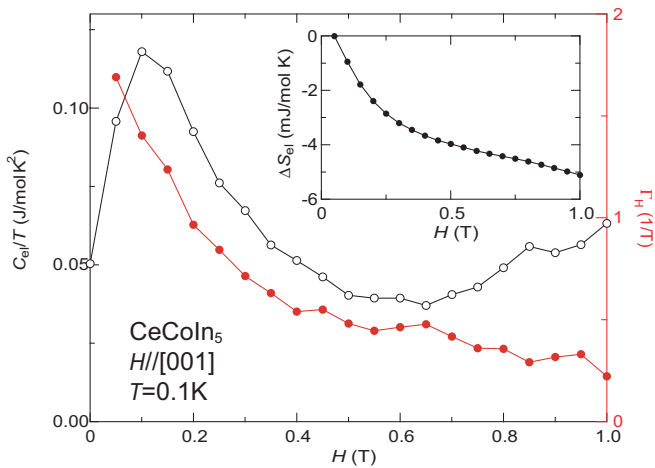}
\caption{(Color online) Electronic specific heat divided by temperature, $C_{el}/T$ (black open circles, left axis), and magnetic Gr\"uneisen ratio, $\Gamma_{\rm H}$ (red solid circles, right axis), of CeCoIn$_5$ as a function of applied field along [001] at 0.1~K. Inset shows increment of isothermal electronic entropy as a function of field at 0.1~K.}
\end{figure}

Our observation of zero-field quantum critical scaling sheds new light on the recent proposals of a field-induced QCP near 4~T (for $H\parallel [001]$), which has been obtained from linear extrapolation of the NFL-FL crossover in thermal expansion, magnetoresistance and Hall effect measurements~\cite{Singh-prl07,Zaum-prl11}. Clearly, the magnetic Gr\"uneisen ratio, which is the most sensitive thermodynamic probe for a field-induced QCP does not diverge in the approach of 4~T but in the approach of zero field (cf. inset of Fig. 2). If we thus exclude a QCP at 4~T, we need to demonstrate, that all the previous measurements would also be compatible with zero-field quantum criticality.  For this purpose, we consider the $T$-$H$ phase diagram shown in Fig. 3, where the color coding represents the size of $\Gamma_{\rm H}/H$ in the normal state, which indicates the entropy accumulation due to quantum criticality~\cite{zhu,GarstM:SigctG}. Previous NFL to FL crossovers are included as circles, whose linear extrapolation would yield $H_c\approx 4$~T~\cite{Singh-prl07,Zaum-prl11}. However, our $T/H^{3/2}$ scaling (Fig. 2) proves that such a linear extrapolation is not justified, since the temperature scale in the critical scaling regime does not depend linearly on $H$ but rather super-linear, $H^{3/2}$. The super-linear crossover between NFL and FL states is indicated by the two dashed lines in the phase diagram, which correspond to the positions of the maxima and inflection points of the magnetic G\"uneisen ratio data of Fig. 2. We note, that the previously determined crossovers all lie in between these two lines, indicating that the these experiments are also compatible with zero-field quantum criticality.
%Skeptics may still argue that the normal state at low temperatures for $H\geq 0$ is far away from zero field. Therefore, signatures of quantum criticality within the SC phase close to zero field would provide additional support of our proposal.

In order to investigate signatures of zero-field quantum criticality on SC quasiparticles, we studied the low-field SC state by combined measurements of $\Gamma_{\rm H}$ and specific heat. Since $\partial S/\partial H=-C_{el}\Gamma_{\rm H}$, we obtain the isothermal field-evolution of the entropy by integration. In our previous work we focused attention near the SC upper critical field and found for $H\parallel$[001] a broad kink in the field dependencies of $\Gamma_{\rm H}$ and $C_{el}$ near 4.4~T\cite{Tokiwa-prl12}. Since this anomaly vanishes upon rotating the field towards the [100] direction, it cannot be related to the isotropic quantum critical scaling. We now concentrate on the behavior close to zero field. Figure 4 displays the measured heat capacity and magnetic Gr\"uneisen ratio at low temperatures, together with the evolution of the entropy (see inset). In nodal superconductors, quasiparticles exist at the gap nodes. Within the Shubnikov phase, the applied magnetic field creates vortices whose cores host additional quasiparticles. We therefore expect an increase of the entropy with increasing field for superconductors and a related negative sign of $\Gamma_{\rm H}$. As shown in Fig. 4, remarkably the magnetic Gr\"uneisen ratio is positive and the entropy {\it decreases} with increasing field in contrary to the expectation for nodal superconductors and multi-band superconductivity~\cite{Seyfarth-prl08}.
%Only at higher fields it recovers the usual negative values, as shown in our previous study~\cite{Tokiwa-prl12}. Thus, the anomalous behavior near zero field likely results from quantum criticality.
% Since the entropy, $S(H)$, is expected to increase monotonically with increasing field due to the increasing number of quasiparticles and $\Gamma_{\rm H}$ is proportional to $-\partial S/\partial H$, $\Gamma_{\rm H}$ is expected to be always negative in a SC state. In the inset of Fig.~3, the measured $\Gamma_{\rm H}$ is shown at low fields inside the SC state. All the measured data of $\Gamma_{\rm H}$ shows an abrupt drop at $T_c$(=2.3K for $H$=0), however it increases anomalously at low temperatures. We ascribe this anomalous increase as SC QPs strongly fluctuating due to the QCP located at zero field. 
Relatedly the isothermal field dependence of the specific heat coefficient $C_{el}/T$ at 0.1~K (cf. Fig. 4) differs from the expected monotonic increase proportional to $\sqrt{H}$ for superconductors with line nodes of the gap~\cite{Volovik,Moler-prl94}. The pronounced reduction of $C_{el}/T$ with increasing field indicates that the quasiparticle mass is strongly enhanced near zero-field. Thus, the peak in the field dependence of $C_{el}/T$ arises from superposition of the usual $\sqrt{H}$ dependence with a quantum critical contribution. Near zero field, the latter vanishes due to the disappearance of vortices. $\Gamma_{\rm H}$ also shows a steep increase towards low fields consistent with zero field quantum criticality. The anomalous decrease of quasiparticle entropy with increasing field (cf. inset of Fig. 4) is thus a consequence of this QCP near $H=0$.
%We note that the thermal conductivity in the SC state of CeCoIn$_5$ also detected anomalous behavior in its field dependence at low fields, qualitatively different from that in multi-band superconductors like MgB$_2$ or PrOs$_4$Sb$_{12}$, and can not be explained by vortex scattering~\cite{Seyfarth-prl08}.
This is compatible with measurements of magnetic penetration depth, which found evidence for nodal quantum criticality at weak magnetic fields~\cite{Hashimoto-pnas13}.

To conclude, a systematic study of the magnetic Gr\"uneisen ratio of CeCoIn$_5$ has revealed quantum critical scaling behavior in the normal state that indicates a critical field very close to zero in contrast to previous claims of a critical field only slightly below the upper critical field of superconductivity. The anomalous field dependence of the quasiparticle entropy within the SC state near zero field further supports this conclusion. It has been discussed in several classes of unconventional superconductors that $T_c$ reaches its maximum near magnetic QCPs~\cite{broun-nphys08,mathur:nature-98,Pfleiderer-RMP09,Ni-prb09,Puban-JPhys11,Lefebvre-prl00,Kusmartseva-prl09}. Since CeCoIn$_5$ exhibits nearly the highest $T_c$=2.3\,K among not only the 115 family, but also all the structural variants of Ce$_n$M$_m$In$_{3n+2m}$ (M; transition metal)~\cite{Thompson-115}, it is not surprising that CeCoIn$_5$ is located at a QCP without tuning any control parameters like composition, pressure of magnetic field. The observed scaling reveals $\nu z=3/2$, consistent with previous thermal Gr\"uneisen parameter studies~\cite{Donath-prl08}, and $\nu d$=0.68. Assuming (quasi) two-dimensional critical fluctuations, this yields an anomalous correlation length exponent of $\nu=1/3$ that characterizes the nature of the QCP in this material.

Stimulating discussions with M. Garst are acknowledged. The work has been supported by the German Science Foundation through FOR 960 (Quantum phase transitions). Work at Los Alamos was performed under the auspices of the U.S. DOE, Office of Basic Energy Sciences, Division of Materials Science and Engineering.

\end{document}